\documentclass[11pt,showpacs,preprint]{revtex4}
\usepackage{graphicx}
\begin{document}

\title{Role of the total isospin {\bf $3/2$} component in 
three-nucleon  reactions
}
\author{H.~Wita{\l}a, J.~Golak, R.~Skibi\'nski, K.~Topolnicki}
\address{M. Smoluchowski Institute of Physics, Jagiellonian
 University, PL-30348 Krak\'ow, Poland}
\author{E.~Epelbaum}
\address{Institut f\"ur Theoretische Physik II, Ruhr-Universit\"at
  Bochum, D-44780 Bochum, Germany}
\author{K.~Hebeler}
\address{Institut f\"ur Kernphysik, Technische Universit\"at 
Darmstadt, D-64289 Darmstadt, Germany\\
Extreme Matter Institute EMMI, GSI Helmholtzzentrum f{\"u}r
Schwerionenforschung GmbH, D-64291 Darmstadt, Germany}
\author{H.~Kamada}
\address{Department of Physics, Faculty of Engineering,
Kyushu Institute of Technology, Kitakyushu 804-8550, Japan}
\author{H.~Krebs}
\address{Institut f\"ur Theoretische Physik II, Ruhr-Universit\"at
  Bochum, D-44780 Bochum, Germany}
\author{U.-G. Mei{\ss}ner}
\address{Helmholtz-Institut f\"ur Strahlen- und
             Kernphysik and Bethe Center for Theoretical Physics, \\
             Universit\"at Bonn,  D--53115 Bonn, Germany\\
             Institute~for~Advanced~Simulation, Institut~f\"{u}r~Kernphysik,
J\"{u}lich~Center~for~Hadron~Physics, and JARA~-~High~Performance~Computing
Forschungszentrum~J\"{u}lich,
D-52425~J\"{u}lich, Germany}
\author{A.~Nogga}
\address{Institut f\"ur Kernphysik, Institute for Advanced Simulation 
and J\"ulich Center for Hadron Physics, Forschungszentrum J\"ulich, 
D-52425 J\"ulich, Germany}
\date{\today}

\begin{abstract}
We discuss the  role of  the 
 three-nucleon isospin $T=3/2$  amplitude  in 
 elastic neutron-deuteron  
scattering and in the deuteron breakup reaction. The contribution 
of this amplitude originates  
from charge-independence breaking  of the nucleon-nucleon  potential 
and is driven by the difference between neutron-neutron   
(proton-proton) and neutron-proton  forces.  
We study the magnitude of that contribution to the elastic scattering 
and breakup observables, taking the locally regularized chiral N$^4$LO 
nucleon-nucleon  potential supplemented by the chiral N$^2$LO 
 three-nucleon force. For comparison we employ also the Av18 nucleon-nucleon 
  potential combined with the Urbana IX three-nucleon force. 
 We find that the isospin $T=3/2$ component is important for the breakup 
reaction and the proper treatment of charge-independence breaking 
  in this case   requires the inclusion of the $^1S_0$ 
state with isospin $T=3/2$. For neutron-deuteron  elastic 
 scattering the $T=3/2$ contributions 
are insignificant and charge-independence breaking 
  can be accounted for by using the effective t-matrix 
generated with the so-called $``2/3-1/3"$ rule. 
\end{abstract}


\maketitle

\section{Introduction}
\label{intro}

Charge-independence breaking (CIB) is well established in the two-nucleon 
 (2N) system in the $^1S_0$ state  as evidenced by the values of the scattering 
lengths $-23.75 \pm 0.01$, $-17.3 \pm 0.8$, and $-18.5 \pm 0.3$~ fm 
 \cite{schori87,teramond87} for 
the neutron-proton (np), proton-proton (pp) 
 (with the Coulomb force subtracted), and neutron-neutron (nn) systems, 
 respectively. 
 That knowledge of CIB is 
incorporated into modern, high precision NN potentials,  as exemplified 
by the standard semi-phenomenological models: 
 Av18 \cite{av18}, CD~Bonn \cite{cdbonn}, or NijmI and 
 NijmII \cite{nijm}, as well as by the chiral 
NN forces~\cite{epel6a,Epelbaum:2008ga,machl6b}.  Treating neutrons and
  protons as identical 
particles  requires 
that nuclear systems are described not only in terms of the momentum and spin 
but also isospin states.  
 The general classification of  the isospin dependence of the NN force is 
given in \cite{henley1979}. 
 The isospin violating 2N forces induce an admixture of 
 the total isospin $T={3/2}$ state 
to the dominant   $T={1/2}$ state in the three-nucleon (3N) system. 
The CIB of the NN interaction thus affects  3N 
 observables. The detailed treatment of the 3N system with CIB NN forces 
in the case of distinguishable or identical particles was formulated 
and described 
in \cite{wit91-cib}. We extend the investigation done in \cite{wit91-cib}
by including a three-nucleon force (3NF). In the calculations performed with 
the standard semi-phenomenological potentials we use 
the  UrbanaIX (UIX) \cite{uIX} 3NF, while 
the chiral N$^2$LO 3N force~\cite{epel2002} is used in addition to 
the recent and most accurate chiral NN interactions \cite{epel1,epel2}.
In this paper, based on such dynamics, we discuss the role of the amplitude 
 with the 
total three-nucleon (3N) isospin $T=3/2$  in elastic neutron-deuteron (nd) 
scattering and in the corresponding breakup reaction. 
In Sec.~\ref{section2} we briefly describe the formalism of 
3N continuum Faddeev 
calculations and the inclusion of CIB.  
 The results are presented in Sec.~ \ref{res}. 
In Sec.~\ref{elastic}  we discuss  
 our results for elastic 
 nd scattering and  in Sec.~ \ref{breakup} describe our findings
  for selected breakup configurations.  
 We summarize and conclude in Sec. ~\ref{summary}.

\section{3N scattering and charge independence breaking}
\label{section2}

Neutron-deuteron scattering with nucleons interacting
through a NN interaction $v_{NN}$ and a 3NF $V_{123}$, is
described in terms of a breakup operator $T$ satisfying the
Faddeev-type integral equation~\cite{wit88,glo96,hub97}
\begin{eqnarray}
T\vert \phi \rangle  &=& t P \vert \phi \rangle +
(1+tG_0)V^{(1)}(1+P)\vert \phi \rangle + t P G_0 T \vert \phi \rangle \cr 
&+& 
(1+tG_0)V^{(1)}(1+P)G_0T \vert \phi \rangle \, .
\label{eq1a}
\end{eqnarray}
The two-nucleon $t$-matrix $t$ is the solution of the
Lippmann-Schwinger equation with the interaction
$v_{NN}$.   $V^{(1)}$ is the part of a 3NF which is 
symmetric under the interchange of nucleons $2$ and $3$: $V_{123}=V^{(1)}(1+P)$.
 The permutation operator $P=P_{12}P_{23} +
P_{13}P_{23}$ is given in terms of the transposition operators,
$P_{ij}$, which interchange nucleons $i$ and $j$.  The incoming state 
$\vert \phi \rangle = \vert \mathbf{q}_0 \rangle \vert \phi_d \rangle$
describes the free relative motion of the neutron and the deuteron 
  with the relative momentum
$\mathbf{q}_0$  and contains the internal deuteron state $\vert \phi_d \rangle$.
Finally, $G_0$ is the resolvent of the three-body center-of-mass kinetic
energy. 
The amplitude for elastic scattering leading to the corresponding
two-body final state $\vert \phi ' \rangle$ is then given by~\cite{glo96,hub97}
\begin{eqnarray}
\langle \phi' \vert U \vert \phi \rangle &=& \langle \phi' 
\vert PG_0^{-1} \vert 
\phi \rangle + 
\langle \phi' \vert PT \vert \phi \rangle + \langle 
\phi'\vert  V^{(1)}(1+P)\vert \phi \rangle  \cr
&+& \langle \phi' \vert V^{(1)}(1+P)G_0T\vert  \phi \rangle,
\label{eq3}
\end{eqnarray}
while for the breakup reaction one has
\begin{eqnarray}
\langle  \phi_0'\vert U_0 \vert \phi \rangle &=&\langle 
 \phi_0'\vert  (1 + P)T\vert
 \phi \rangle ,
\label{eq3_br}
\end{eqnarray}
where $\vert \phi_0' \rangle$ is the free three-body breakup channel state. 

Solving Eq.(\ref{eq1a}) in the momentum-space partial wave basis, defined by 
the magnitudes of the 
3N Jacobi momenta $p$ and $q$ together with the angular momenta and isospin 
quantum numbers $\alpha$ ($\beta$),     
 is performed by projecting Eq.~(\ref{eq1a}) onto two types  of 
basis states:
\begin{eqnarray}
\vert p q \alpha \rangle \equiv \vert p q ~{\rm angular~momenta} \rangle 
\vert (t {1\over{2}})T={1\over {2}} M_T \rangle \, , ~ (t=0,1) \, ,
\label{eq4}
\end{eqnarray}
and 
\begin{eqnarray}
\vert p q \beta \rangle \equiv \vert p q ~{\rm angular~momenta} \rangle 
\vert (t {1\over{2}})T={3\over {2}} M_T \rangle \, ,  ~ (t=1) \, .
\label{eq5}
\end{eqnarray}
Assuming charge conservation and employing the notation 
where the neutron (proton) isospin projection is  
${1\over{2}}$ ($-{1\over{2}}$),  the 2N t-operator in 
the three-particle isospin 
space can be decomposed for the nd system as~\cite{wit91-cib}:
\begin{eqnarray}
\langle (t{1\over{2}}) T M_T={1\over{2}} \vert t \vert  
(t'{1\over{2}}) T' M_{T'}={1\over{2}} \rangle &=& 
\delta_{tt'} \delta_{TT'}\delta_{T1/2} [ \delta_{t0} t^{t=0}_{np} +
\delta_{t1} ( {2 \over{3}} t^{t=1}_{nn} +  {1 \over{3}} t^{t=1}_{np} ) ] \cr 
&+& \delta_{tt'} \delta_{t1} (1-\delta_{TT'}) 
{\sqrt{2} \over {3}} ( t^{t=1}_{nn} -  t^{t=1}_{np} ) \cr 
&+&\delta_{tt'} \delta_{t1} \delta_{TT'}\delta_{T3/2} 
 ( {1 \over{3}} t^{t=1}_{nn} +  {2 \over{3}} t^{t=1}_{np} ) ~,
\label{eq6}
\end{eqnarray}
where $t_{nn}$ and $t_{np}$ are solutions of the Lippman-Schwinger equations 
driven by the $v_{nn}$ and $v_{np}$ potentials, respectively.

As a result of solving Eq.~(\ref{eq1a}) one gets the amplitudes 
$\langle  p q \alpha \vert T \vert \phi \rangle$ and 
  $\langle p q \beta \vert T \vert \phi \rangle$, 
which fulfill the following set of coupled integral equations:
\begin{eqnarray}
\langle p q \alpha \vert T\vert \phi \rangle  &=& 
 \sum_{\alpha'} \int_{p'q'} \langle p q \alpha \vert t \vert p' q' \alpha'\rangle
\langle p' q' \alpha' \vert P \vert \phi \rangle \cr 
&+&
\sum_{\alpha'} \int_{p'q'} \langle p q \alpha \vert V^{(1)}\vert p' q' 
\alpha'\rangle 
\langle p' q' \alpha' \vert (1+P)\vert \phi \rangle \cr
&+& 
\sum_{\alpha'}\int_{p'q'} \langle p q \alpha \vert t \vert p' q' \alpha'\rangle 
\langle p' q' \alpha' \vert G_0 V^{(1)} (1+P)\vert \phi \rangle \cr
&+&
\sum_{\alpha'}\int_{p'q'} \langle pq \alpha \vert t \vert p' q' \alpha'\rangle 
\langle p' q' \alpha' \vert P G_0 T \vert \phi \rangle \cr 
&+&
\sum_{\beta'} \int_{p'q'} \langle pq \alpha \vert t \vert p' q' \beta'\rangle 
\langle p' q' \beta' \vert P G_0 T \vert \phi \rangle \cr 
&+& 
\sum_{\alpha'} \int_{p'q'} \langle p q \alpha \vert V^{(1)}\vert p' q' 
\alpha'\rangle 
\langle p' q' \alpha' \vert (1+P)G_0T \vert \phi \rangle \cr
&+&
\sum_{\alpha'} \int_{p'q'}   \sum_{\alpha''} \int_{p''q''}  
\langle p q \alpha \vert t \vert p' q' \alpha'\rangle 
\langle p' q' \alpha' \vert G_0V^{(1)} \vert p''q''\alpha''\rangle \cr
&\times& \langle p''q''\alpha''\vert (1+P)G_0T \vert \phi \rangle \cr
&+&
\sum_{\beta'} \int_{p'q'} \sum_{\beta''} \int_{p''q''} 
\langle p q \alpha \vert t \vert p' q' \beta'\rangle 
\langle p' q' \beta' \vert G_0V^{(1)} \vert p''q''\beta''\rangle \cr
&\times& \langle p''q''\beta''\vert (1+P)G_0T \vert \phi \rangle \cr
\langle p q \beta \vert T\vert \phi \rangle  &=& 
 \sum_{\alpha'} \int_{p'q'} \langle p q \beta \vert t \vert p' q' \alpha'\rangle
\langle p' q' \alpha' \vert P \vert \phi \rangle \cr 
&+&
\sum_{\alpha'} \int_{p'q'} \langle p q \beta \vert t \vert p' q' \alpha'\rangle 
\langle p' q' \alpha' \vert G_0 V^{(1)} (1+P)\vert \phi \rangle \cr
&+&
\sum_{\alpha'} \int_{p'q'} \langle p q \beta \vert t \vert p' q' \alpha'\rangle 
\langle p' q' \alpha' \vert P G_0 T \vert \phi \rangle \cr 
&+&
\sum_{\beta'} \int_{p'q'} \langle p q \beta \vert t \vert p' q' \beta'\rangle 
\langle p' q' \beta' \vert P G_0 T \vert \phi \rangle \cr 
&+& 
\sum_{\beta'} \int_{p'q'} \langle p q \beta \vert V^{(1)}\vert p' q' \beta'\rangle 
\langle p' q' \beta' \vert (1+P)G_0T \vert \phi \rangle \cr
&+&
\sum_{\alpha'} \int_{p'q'} \sum_{\alpha''} \int_{p''q''} 
  \langle p q \beta \vert t \vert p' q' \alpha'\rangle 
\langle p' q' \alpha' \vert G_0V^{(1)} \vert p'' q''\alpha''\rangle \cr
&\times& \langle p'' q''\alpha''\vert 
 (1+P)G_0T \vert \phi \rangle \cr
&+&
\sum_{\beta'} \int_{p'q'} \sum_{\beta''} \int_{p''q''} 
\langle p q \beta \vert t \vert p' q' \beta'\rangle 
\langle p' q' \beta' \vert G_0V^{(1)} \vert p''q''\beta''\rangle    \cr
&\times& \langle p'' q''\beta''\vert 
 (1+P) G_0T \vert \phi \rangle ~.
\label{eq7}
\end{eqnarray}
The form of the couplings in Eq.~(\ref{eq7}) follows from the fact that 
the incoming neutron-deuteron state $\vert \phi \rangle$ is a
total isospin $T=1/2$ state, the permutation operator $P$ is diagonal 
in the total isospin,  
and  the 3NF is assumed to conserve the total isospin  $T$~\cite{epel2005}. 

From Eq.~(\ref{eq7}) it is clear that only when the nn and np 
interactions differ in the same orbital and spin angular-momentum states 
 with the 2N 
subsystem isospin $t=1$ (CIB), then the amplitudes  
 $\langle p q \beta \vert T \vert \phi \rangle$ will be nonzero. 
 In addition, their magnitude is driven  by the strength of the CIB 
 as given by the difference of the corresponding t-matrices 
${\sqrt{2} \over {3}} ( t^{t=1}_{nn} -  t^{t=1}_{np} )$ in Eq.~(\ref{eq6}). 
In such a case, not only the magnitude of CIB decides about the importance of 
the  $\langle p q \beta \vert T \vert \phi \rangle$ contributions, but also the 
 isospin $T=3/2$ 3NF matrix elements, which  participate in generating 
the   $\langle p q \beta \vert T \vert \phi \rangle$  amplitudes. 
It is the set of equations Eq.~(\ref{eq7}) which we solve when we differentiate 
between nn and np interactions and include both $T=1/2$ and $T=3/2$ 3NF 
matrix elements.

In the case when the neglect of the $T=3/2$ amplitudes 
 $\langle p q \beta \vert T \vert \phi \rangle$ is 
justified, the CIB can be  taken care of by using
 the effective two-body t-matrix generated with the 
 so-called $``2/3-1/3"$ rule, 
$t_{\rm eff}={2 \over{3}} t^{t=1}_{nn} +  {1 \over{3}} t^{t=1}_{np}$, 
 see Eq.(\ref{eq6}),  
in the 2N subsystem isospin $t=1$ states 
 and restricting the treatment only to the amplitudes  
$\langle p q \alpha \vert T \vert \phi \rangle$. 

Since the final state $\vert \phi' \rangle$ in elastic nd scattering also 
 has the 
total 3N isospin $T=1/2$, the amplitudes 
  $\langle p q \beta \vert T \vert \phi \rangle$ do 
not contribute directly to this reaction. The $T=3/2$ admixture enters
in this case through a modification of the $T=1/2$ amplitudes 
 $\langle p q \alpha \vert T \vert \phi \rangle$ induced by the couplings given 
 in  Eq.~(\ref{eq7}). 
Contrary to that, for the nd breakup reaction both $T=1/2$ 
 ($\langle p q \alpha \vert T \vert \phi \rangle$) and $T=3/2$  
 ($\langle p q \beta \vert T \vert \phi \rangle$) amplitudes contribute.

\section{Results}
\label{res}

In order to check the importance of the isospin $T=3/2$ contributions we solved 
the 3N Faddeev equations Eq.~(\ref{eq7}) for four values of the 
 incoming neutron 
laboratory energy: $E_{lab}=13$, $65$, $135$, and $250$~MeV. 
 As a  NN potential we took the semi-locally regularized
N$^4$LO chiral potential of ref.~\cite{epel1,epel2,binder2016} with 
the regulator $R=0.9$~fm, alone or 
combined with the chiral 
N$^2$LO 3NF~\cite{epel2002,heb1}, regularized with 
the same regulator. We additionally regularized   
matrix elements of that 3NF by multiplying it with a nonlocal regulator 
$f(p,q)=\exp\{-(p^2+{3\over{4}}q^2)^3/{\Lambda}^6\}$  
with large cut-off value $\Lambda=1000$~MeV. 
 This additional regulator is applied to the 3NF matrix elements only 
for technical reasons. The practical calculation of the local 3NFs involve 
the evaluation of convolution integrals whose calculation becomes 
numerically unstable at very large momenta. The value of the cutoff scale 
 $\Lambda$ is chosen sufficiently large so that low-energy physics is not 
affected by this additional regulator. In fact, we have checked 
explicitly that the effects of this regulator for the chosen cutoff 
value are negligible in three-body bound state and scattering calculations.
 As a nn force we took the pp 
version of that particular NN interaction (with the Coulomb force subtracted). 
 The low-energy constants  
of the contact interactions in that 3NF were adjusted to the 
 triton binding energy 
and we used  $c_D=6.0$ for the one-pion exchange contribution 
and $c_E=-1.0943$ for the 3N contact term (we are using the notation of 
Ref.~\cite{epel2002} with $\Lambda=700$~MeV).  That specific choice of 
the $c_D$ and $c_E$ values does not only reproduce the experimental 
triton binding energy when that N$^4$LO NN and N$^2$LO 3NF are combined 
but also provides quite a good description of the nucleon-deuteron 
 elastic scattering 
cross section data  at higher energies. In order to provide 
convergent predictions we solved Eq.~(\ref{eq7}) 
taking into account all partial wave states with the total 2N angular 
momenta up to $j_{max}=5$ and 3N total angular momenta up to $J_{max}=25/2$. 
The 3NF was included up to $J_{max}=7/2$. 

We present results for that particular combination of the chiral 
 NN and 3N forces (of course, in the future a more consistent set of 2N 
 and 3N forces
needs to be employed, once the corresponding 3NFs are available) . 
However, we checked using the example of 
the  Av18 and Urbana IX 3NF combination that the conclusions remain 
unchanged when instead of the chiral forces so-called high 
 precision realistic forces are used. 
 
\begin{figure}
\begin{center}
\begin{tabular}{c}
\resizebox{120mm}{!}{\includegraphics{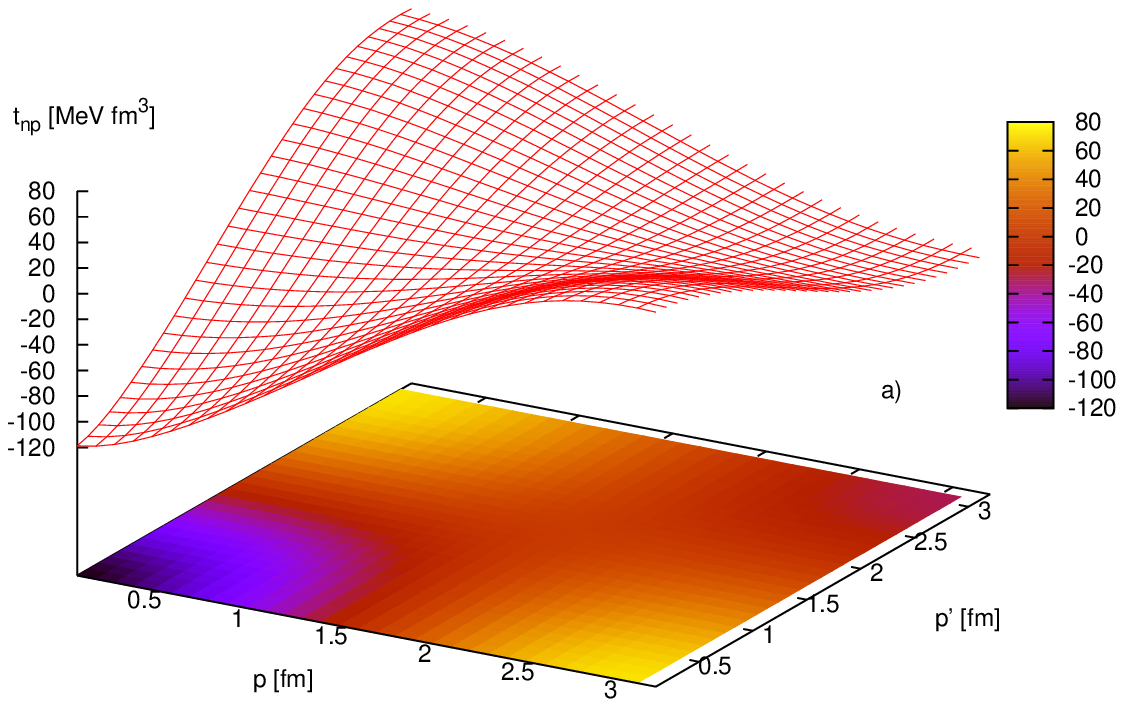}} \\
\resizebox{120mm}{!}{\includegraphics{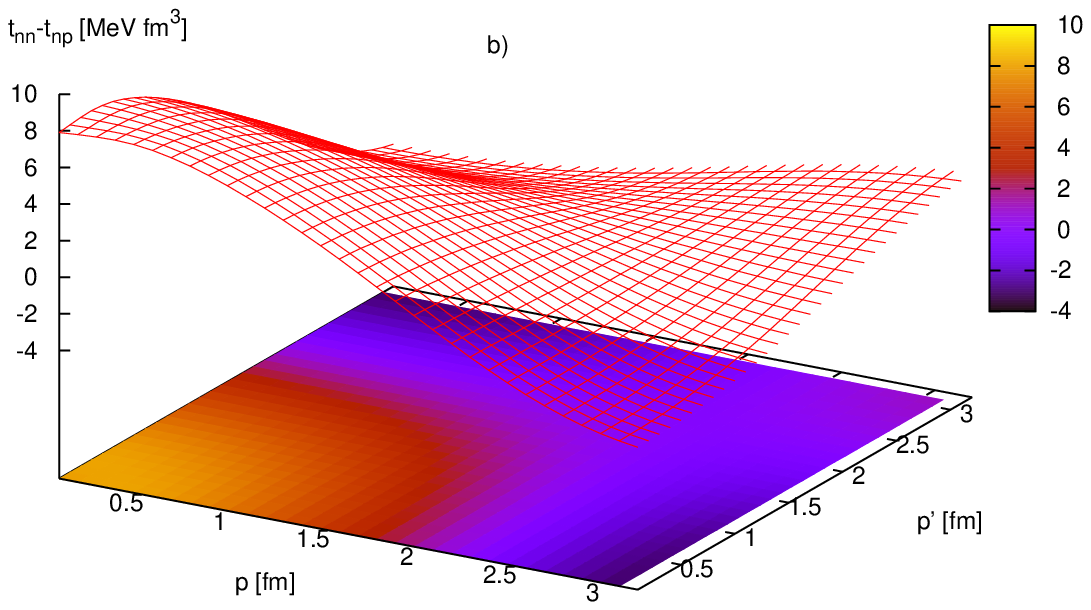}} \\
\end{tabular}
\caption{
(color online) 
The np t-matrix $t_{np}(p,p';E-{3\over{4}}q^2)$ (a) and the difference of 
the nn and np t-matrices (b) at incoming neutron laboratory 
energy $E_{lab}=13$~MeV 
for the $^1S_0$ partial wave, as a function of the relative 
NN momenta $p$ and $p'$ at a particular value of the 
spectator nucleon momentum $q=0.528$~fm$^{-1}$ at which the 2N subsystem 
 energy is equal to the   
 binding energy of the deuteron $E_d$: $E-{3\over{4}}q^2=E_d$.}
\label{fig1}
\end{center}
\end{figure}

\begin{figure}
\begin{center}
\begin{tabular}{c}
\resizebox{120mm}{!}{\includegraphics{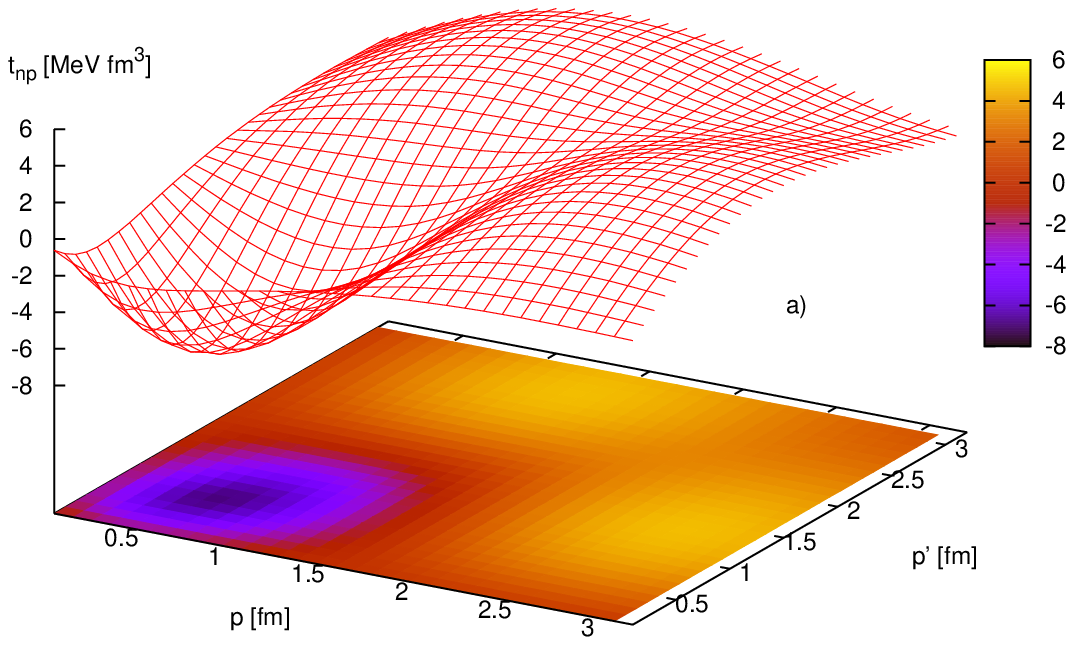}} \\
\resizebox{120mm}{!}{\includegraphics{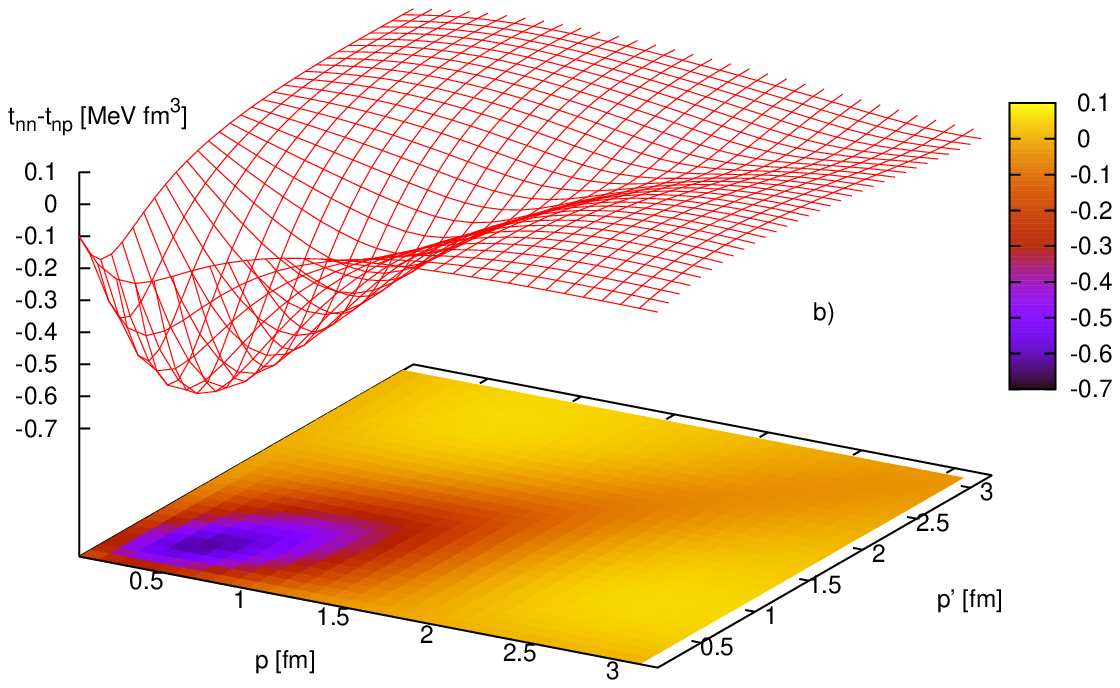}} \\
\end{tabular}
\caption{
(color online) 
The same as in Fig.~\ref{fig1} but for the $^3P_0$ partial wave.}
\label{fig2}
\end{center}
\end{figure}

Since CIB effects are driven by the difference between np and nn t-matrices we
display in Figs.~\ref{fig1} and \ref{fig2} 
 for the $^1S_0$ and  $^3P_0$ NN partial waves, respectively, the np t-matrix 
$t_{np}(p,p';E-{3\over{4}}q^2)$ (a) and 
the difference $t_{np}(p,p';E-{3\over{4}}q^2)-t_{nn}(p,p';E-{3\over{4}}q^2)$ (b),
 at the laboratory energy of the 
incoming neutron  $E_{lab}=13$~MeV. They are displayed as a function of 
the NN relative momenta 
$p$ and $p'$ for a chosen value of the spectator nucleon momentum 
$q=0.528$~fm$^{-1}$ at which the 2N subsystem energy is equal to the   
 binding energy of the deuteron $E_d$: $E-{3\over{4}}q^2=E_d$. 
 The behaviour of the $t_{np}$ 
as well as of the difference  $t_{np}-t_{nn}$ is similar at other energies. 
It is interesting to note that the difference between the np and nn 
t-matrices ranges  up to $\approx 10~\%$.

In Tables~\ref{table1},~\ref{table2},~and~\ref{table3} we show at the chosen 
energies  the total cross section for the nd interaction, the total 
nd elastic scattering cross section, and the total nd breakup 
cross section, respectively, calculated with different underlying dynamics 
based on the chiral N$^4$LO NN or/and N$^2$LO 3NF force.  
Namely, the results in column~2 of those tables (no CIB, 
 $V_{123}=0$, $^1S_0$ np) 
were obtained with the NN force only, assuming no CIB and using in all $t=1$ 
partial waves the effective t-matrix $t_{\rm eff}=(2/3)t_{nn}+(1/3)t_{np}$, with 
 the exception of the $^1S_0$ partial wave, where only the np force  was taken.
  
In column~3 instead of the np $^1S_0$ NN force a nn one was taken 
(no CIB, $V_{123}=0$, $^1S_0$ nn). In column~4 in all $t=1$ partial waves 
(including also $^1S_0$ one) only $T=1/2$ was taken into account and 
the effective t-matrix $t_{\rm eff}=(2/3)t_{nn}+(1/3)t_{np}$
 was used (no CIB, $V_{123}=0$, $t_{\rm eff}$). In column~5 the NN interaction of 
column~4 was combined with 3NF (no CIB, $V_{123}$, $t_{\rm eff}$). 
In column~6 a proper treatment of the CIB in the $^1S_0$ partial wave 
was performed by taking in that partial wave both np and nn interactions 
and keeping in addition to the total isospin $T=1/2$  also $T=3/2$. 
In all other $t=1$ states the effective t-matrix $t_{\rm eff}$ was used and 
only states with $T=1/2$ were kept. No 3NF was 
 allowed ($^1S_0$ CIB, $V_{123}=0$). 
Results, when in addition also the 3NF was active, are shown in 
column~7 ($^1S_0$ CIB, $V_{123}$). The proper treatment of CIB in all states 
with $t=1$, when both np and nn interactions were used and 
both $T=1/2$ and $T=3/2$ states were kept, are shown in column~8 and 9 
for the cases when NN interactions were used alone (CIB, $V_{123}=0$) and 
combined with 3NF (CIB, $V_{123}$), respectively. 

From Tables~\ref{table1},~\ref{table2},~and~\ref{table3} it is clear that 
it is sufficient to use the effective t-matrix $t_{\rm eff}$ and to 
 neglect $T=3/2$ 
states completely to account exactly 
for CIB effects in all three total cross sections. That is true in both cases, 
when the 3NF is present or absent. In the case when the 3NF is not  
 included, the total cross 
sections for the nd interaction, for elastic scattering, and for breakup, 
 depend 
slightly on the $^1S_0$ t-matrix used in the calculations. Changing it 
from $t_{np}$ to $t_{nn}$ leads to differences of the cross section values 
 up to $\approx 2\%$ (columns~2 and 4). Using in the $^1S_0$ channel the 
effective t-matrix $t_{\rm eff}=(2/3)t_{nn}+(1/3)t_{np}$ accounts for all CIB 
effects exactly, without the necessity to introduce the total 
isospin $T=3/2$ components 
in any of the $t=1$ partial wave states. Namely, the exact treatment 
 of CIB by using in all 
$t=1$ states the $t_{np}$ and $t_{nn}$ t-matrices and both $T=1/2$ and $T=3/2$ 
 partial wave states (column~8) gives the same value for all three total 
cross sections. Also restricting  the exact treatment of CIB to 
 the $^1S_0$ state 
only (column~6) provides the same values for the total cross sections. It shows 
that contribution of $T=3/2$ states, as far as the total cross sections are 
concerned, can be neglected and all CIB effects  properly taken into 
account by restricting to the total isospin $T=1/2$ states only and using in 
all $t=1$ channels the effective t-matrix generated according 
to the $``2/3-1/3"$ rule.
 
The same is true when 3NF is included. In this case,  
  in all three total cross sections, 
 clear effects increasing with energy  are seen. But again, all three 
treatments of CIB  yield the same numbers (columns 5, 7, and 9). Since in 
the cases when $T=3/2$ states were included (columns 7 and 9) also the 
corresponding $T=3/2$ matrix elements of a 3NF were used, we conclude that 
their influence on the total cross sections is negligible.

\subsection{Elastic scattering}
\label{elastic}

\begin{figure}
\includegraphics[scale=0.65,clip=true]{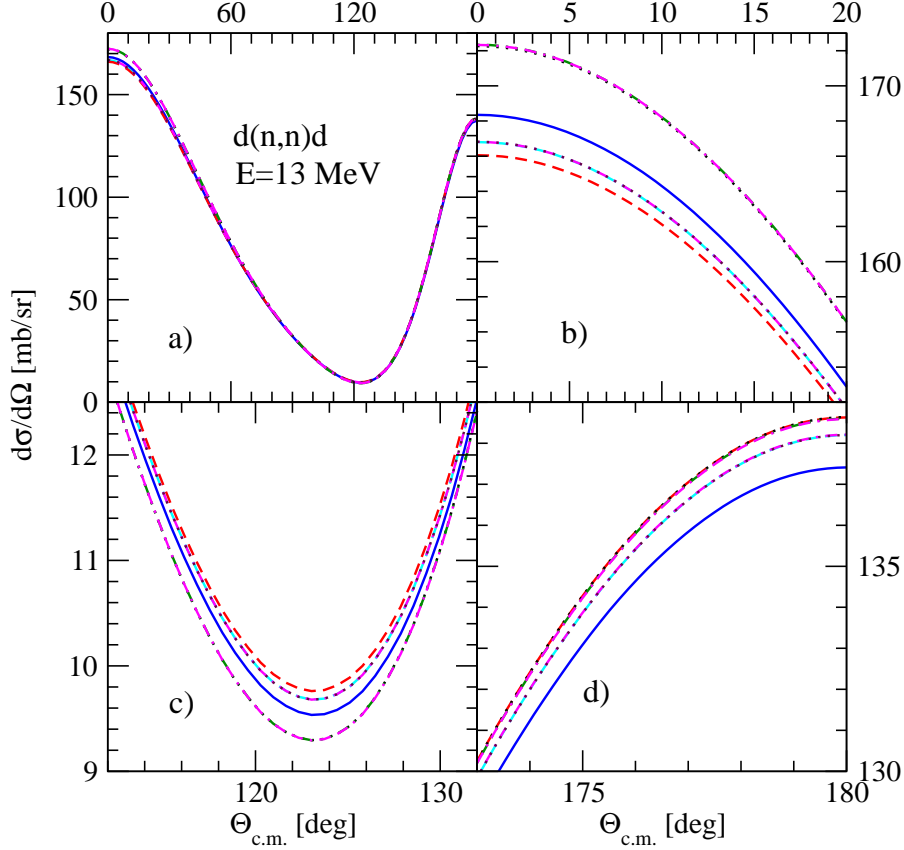}
\caption{
(color online) 
The  nd elastic scattering cross section at $13$~MeV 
 of the incoming neutron laboratory energy. In a) the full angular distribution 
is shown while in b), c) and d) the  forward, intermediate and backward regions 
of angles are displayed. Different lines are predictions obtained with the
locally regularized (regulator $R=0.9$~fm) N$^4$LO NN potential alone or 
combined with the locally regularized N$^2$LO 3NF force for 
different underlying dynamics:  dashed (cyan) line - NN potential alone and 
 in all $t=1$ states effective t-matrix $t_{\rm eff}$ was used, 
solid (blue) line - NN potential alone, in the $^1S_0$ state np force  and 
 in all other $t=1$ states the effective t-matrix $t_{\rm eff}$ was used, 
dashed (red) line - NN potential alone, in the $^1S_0$ state the nn force  and 
 in all other $t=1$ states the effective t-matrix $t_{\rm eff}$ was used,  
dotted (black) line - NN potential combined with 3NF and 
 in all $t=1$ states the effective t-matrix $t_{\rm eff}$ was used,  
 dashed-dotted (magenta) line - NN potential alone, in the $^1S_0$ channels 
 np and nn forces with isospin $T=3/2$ included   and 
 in all other $t=1$ states the effective t-matrix $t_{\rm eff}$ was used, 
 dotted (maroon) line - NN potential alone, in  all  $t=1$ states 
 np and nn forces used and  isospin $T=3/2$ included, 
 dashed-double-dotted (green) line - NN potential combined with 3NF, 
 in $^1S_0$ states  the 
 np and nn forces used and  isospin $T=3/2$ included   and 
 in all other $t=1$ states the effective t-matrix $t_{\rm eff}$ was used, 
  dotted-double-dashed (magenta) line - NN potential combined with 3NF, 
 in  all  $t=1$ states  the np and nn forces used and  isospin $T=3/2$ 
states included.
}
\label{fig3}
\end{figure}

\begin{figure}
\includegraphics[scale=0.65,clip=true]{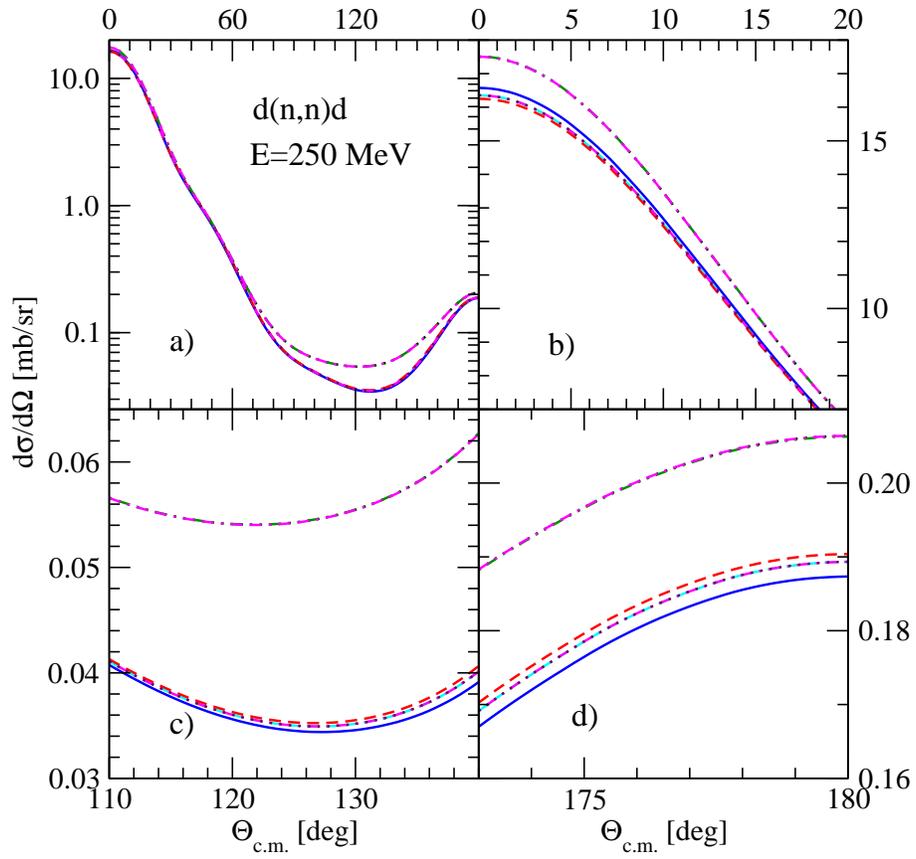}
\caption{
(color online) 
The  same as in Fig.~\ref{fig3} but at  $250$~MeV.
}
\label{fig4}
\end{figure}

In Figs.~\ref{fig3} and \ref{fig4} we display results for the nd elastic 
scattering 
angular distributions  obtained using different assumptions about the 
  underlying 
dynamics and treatment of CIB. As for the total cross sections, in case when 
 the 3NF is inactive, the three treatments of the  CIB, namely 
 using $t_{\rm eff}$ and no 
$T=3/2$ states, $T=3/2$ in the $^1S_0$ state with the $t_{np}$ and $t_{nn}$ 
t-matrices in that state, and $T=3/2$ in all $t=1$ states with 
the corresponding np and nn t-matrices, provide the same elastic 
scattering cross sections (dashed (cyan), dashed-dotted (magenta), and 
dotted (maroon) lines, respectively). These lines overlap  in 
Figs.~\ref{fig3} and \ref{fig4} and the results are  displayed in more detail
 in b), c), and d) for particular ranges of angles). 
 We checked that  also for 
all nd elastic scattering spin observables, encompassing the neutron (vector) 
and deuteron (vector and tensor) analyzing powers, the spin correlation  
 as well as spin transfer coefficients, the above three approaches lead to 
the same results. Thus again the contribution of all $T=3/2$ states can 
be neglected and CIB in elastic nd scattering treated exactly by 
restricting only to $T=1/2$ states and  using in 
all $t=1$ states the $``2/3-1/3"$ rule to generate from $t_{np}$ and $t_{nn}$ 
the effective $t_{\rm eff}$ t-matrix. 

Restricting to $t_{np}$ or $t_{nn}$ t-matrices in the $^1S_0$ channel and 
 neglecting all $T=3/2$ states changes the 
elastic scattering cross sections and spin observables by up to 
$\approx 1\%$ (see the solid (blue) and dashed (red) lines, respectively, 
 in Figs.~\ref{fig3} and \ref{fig4}). 

Adding the 3NF changes the elastic scattering cross section. The 3NF effects
grow with the projectile energy and are especially large in the region of 
intermediate and backward angles. But again the three approaches to CIB 
 provide the same 
cross sections and spin observables (they are displayed for cross sections in 
 Figs.~\ref{fig3} and \ref{fig4} by overlapping lines: dotted (black), 
 dashed-double-dotted (green), and dotted-double-dashed (magenta)). That again
 supports the conclusion that $T=3/2$ states can be neglected together with 
 the $T=3/2$ 3NF matrix elements for all  nd elastic scattering observables.

\subsection{Breakup reaction}
\label{breakup}

In the final breakup state of three free nucleons both $T=1/2$ and $T=3/2$ 
total isospin components are allowed. Thus one would expect that here the  
influence of the $T=3/2$ components will be better visible than in  elastic 
scattering. 

\begin{figure}
\includegraphics[scale=0.65,clip=true]{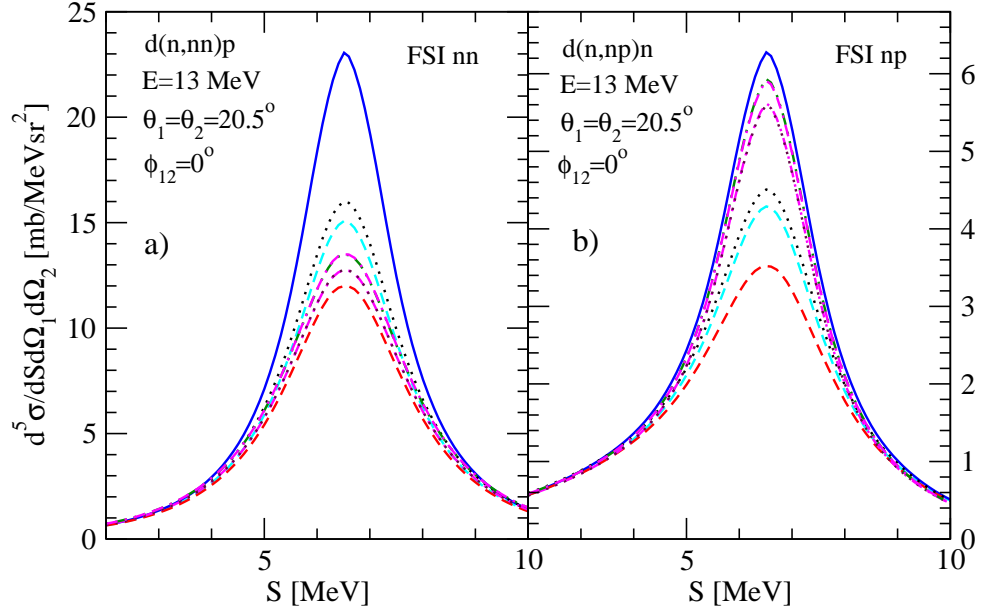}
\caption{
(color online) 
The  nd complete breakup d(n,nn)p 
 cross section $d^5\sigma/d\Omega_1d\Omega_2dS$ 
for the nn (a) and np (b) FSI configuration at $13$~MeV 
 of the incoming neutron laboratory energy and laboratory angles of 
detected outgoing 
nucleons $\theta_1=\theta_2=20.5^o$ and $\phi_{12}=0^o$, as a function of 
the arc-length of the S-curve. For the description of lines see Fig.~\ref{fig3}.
}
\label{fig5}
\end{figure}

\begin{figure}
\includegraphics[scale=0.65,clip=true]{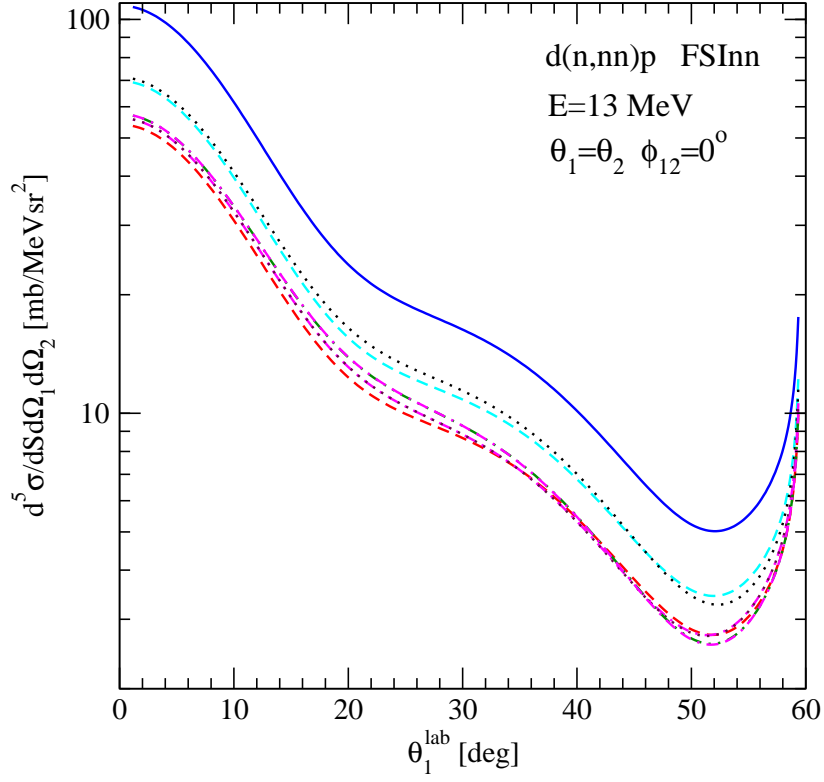}
\caption{
(color online) 
The  nd complete breakup d(n,nn)p 
 cross section $d^5\sigma/d\Omega_1d\Omega_2dS$ calculated exactly at 
 the neutron-neutron final state interaction  condition (maximum 
of the cross section along the S-curve) at $13$~MeV 
 of the incoming neutron laboratory energy as a function of the laboratory 
 production angle of the  outgoing final-state-interacting 
neutrons $\theta_1^{lab}=\theta_2^{lab}$ and $\phi_{12}=0^o$. 
 For the description of lines see Fig.~\ref{fig3}.
}
\label{fig6}
\end{figure}

\begin{figure}
\includegraphics[scale=0.65,clip=true]{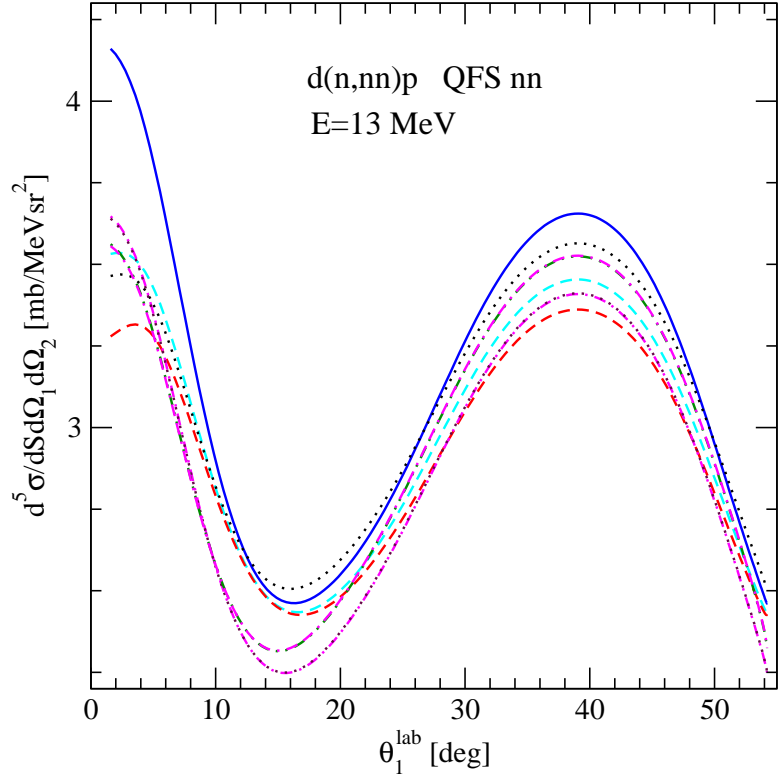}
\caption{
(color online) 
The  nd complete breakup d(n,nn)p 
 cross section $d^5\sigma/d\Omega_1d\Omega_2dS$ calculated exactly at 
 the neutron-neutron quasi-free-scattering condition (maximum 
of the cross section along the S-curve at $E_3^{lab}=0$ and $\phi_{12}=180^o$) 
at $13$~MeV 
 of the incoming neutron laboratory energy as a function of the laboratory 
 angle  of the outgoing 
neutron 1. 
 For the description of lines see Fig.~\ref{fig3}.
}
\label{fig7}
\end{figure}

\begin{figure}
\includegraphics[scale=0.65,clip=true]{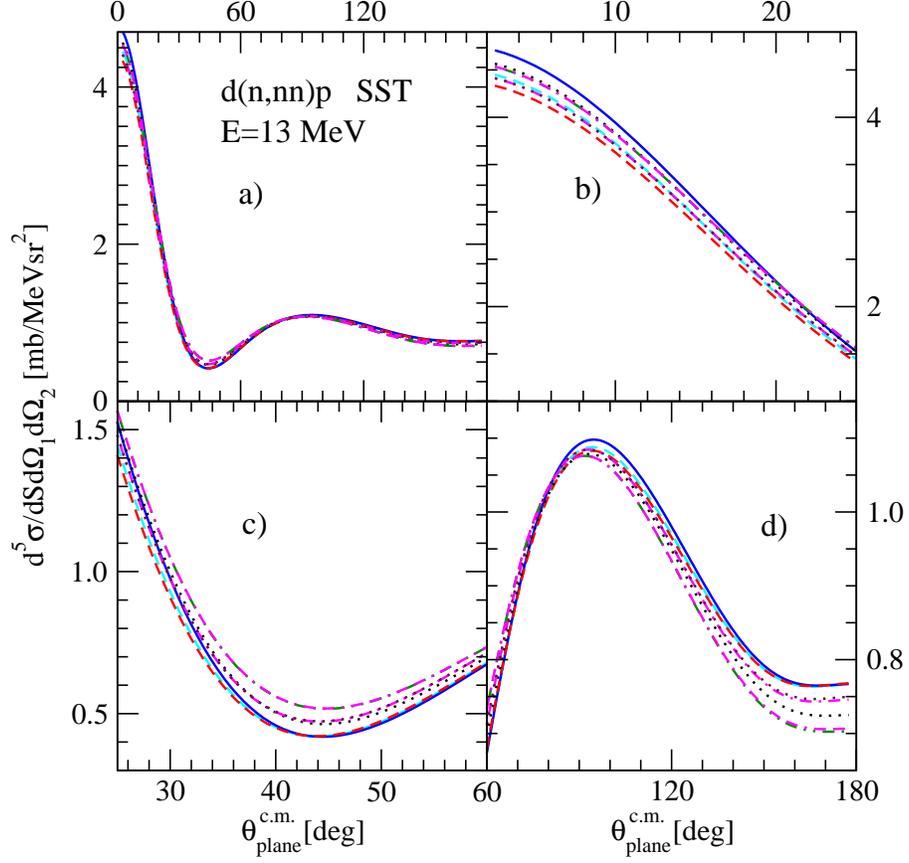}
\caption{
(color online) 
The  nd complete breakup d(n,nn)p 
 cross section $d^5\sigma/d\Omega_1d\Omega_2dS$ exactly at 
 the symmetrical-space-star condition 
 (in the 3N c.m. system the momenta of three outgoing 
nucleons are equal and form a symmetric star in a plane inclined at an 
 angle  $\theta_{\rm plane}^{\rm{c.m.}}$ with respect to 
 the incoming neutron momentum) 
 at $13$~MeV 
 of the incoming neutron laboratory energy, as a function of the  angle 
 $\theta_{\rm plane}^{\rm{c.m.}}$. 
 For the description of lines see Fig.~\ref{fig3}.
}
\label{fig8}
\end{figure}

We show in Figs.~\ref{fig5}-\ref{fig8} that this indeed is the case for the 
 example of 
three kinematically complete breakup configurations: final-state-interaction 
(FSI), quasi-free-scattering (QFS), and symmetrical-space-star (SST).

In the FSI configuration under the exact FSI condition, the two outgoing 
 nucleons have 
equal momenta. Their strong interaction in the $^1S_0$ state leads to a 
characteristic cross section maximum occurring at the exact FSI condition,  
 the magnitude of which is sensitive to the $^1S_0$ scattering length. 
 Since largest CIB effects are seen in the difference between np and nn (pp)  
$^1S_0$ scattering lengths, the region of the FSI peak should reveal 
largest CIB effects. That is clearly demonstrated in 
Fig.~\ref{fig5}a for the nn and in Fig.~\ref{fig5}b for the  
 np FSI configuration, where 
the solid (blue) and dashed (red) lines display FSI cross sections obtained 
with the $^1S_0$ np and nn t-matrices, respectively, using in all other $t=1$ 
states the effective t-matrix $t_{\rm eff}$. Only $T=1/2$ states were used 
and the 3NF was omitted. Using also in the $^1S_0$ state the effective 
t-matrix leads to the dashed (cyan) line, which changes to the 
 dotted (black) line 
when, keeping the rest unchanged, the 3NF is also included. Most 
interesting is the effect of treating the CIB exactly in the state $^1S_0$ by 
including the $T=3/2$ component and using both $t_{np}$ and $t_{nn}$ t-matrices: 
dashed-dotted (magenta) and dashed-double-dotted (green) lines in case when 
3NF is omitted and included, respectively. As expected, the inclusion 
 of the isospin 
$T=3/2$ in the $^1S_0$ state brings the predictions, in the case when 3NF 
is omitted, close to the nn prediction 
(dashed (red) line in Fig.~\ref{fig5}a) for the nn FSI, 
and close to the np prediction 
(solid (blue) line in Fig.~\ref{fig5}b) for the np FSI. However, a significant 
difference exists between that result and the pure nn or np ones as well as 
when compared to  
results obtained with the effective t-matrix $t_{\rm eff}$ (dashed (cyan) line in 
Fig.~\ref{fig5}a and \ref{fig5}b). This shows that the proper treatment of CIB 
in the FSI configurations of the nd breakup requires the inclusion of the 
total isospin 
$T=3/2$ component and using both $t_{np}$ and $t_{nn}$ 
t-matrices   in the $^1S_0$ state.  
This is also sufficient for the exact treatment of CIB as shown 
by the results of the full CIB treatment, where in all $t=1$ partial waves also 
 isospin $T=3/2$ states are taken into account and 
corresponding  $t_{np}$ and $t_{nn}$ t-matrices are used, as shown 
in Figs.~\ref{fig5}a and \ref{fig5}b  
by the dotted (maroon) line for the NN 
 interaction acting alone and dotted-double-dashed 
(magenta) line when combined with the 3NF, respectively. 
 These lines overlap with the
lines corresponding to the case when $T=3/2$ is included for the 
state $^1S_0$. It is interesting to note that the $T=3/2$ component in the 
$^1S_0$ state is important and provides FSI cross sections which are different 
from the results obtained with particular NN $^1S_0$ interactions only 
(np for np FSI and nn for nn FSI). It proves the importance of 
including $T=3/2$ in the $^1S_0$ state and shows that both np and nn 
interactions 
have to be employed when 
the FSI peaks are analyzed to extract the value of the corresponding 
scattering length.

In order to see how the  magnitude of the effects induced by 
 the $T=3/2$ $^1S_0$ component depends on 
the particular FSI configuration, we present in Fig.~\ref{fig6} the
cross section in the maximum of the FSI peak 
 as a function of the laboratory production 
 angle of the final-state 
interacting pair. Again it is clearly seen that restricting to $t_{\rm eff}$ only 
and neglecting $T=3/2$ components is insufficient to include all CIB effects. 
Inclusion of  $T=3/2$ component only in the $^1S_0$ state is, however, 
  sufficient to 
 fully account  for the CIB effects. The importance of that component 
depends on the production angle. At the angles in the region around 
$\approx 45^{o}$ the contribution of that component is tiny but becomes 
significant at smaller and larger production angles.

For the QFS and SST configurations the picture is similar. Again in order 
 to fully 
account for the CIB effects it is necessary and sufficient to include the total 
 isospin $T=3/2$ component in the $^1S_0$ state. 
 We exemplify this in Fig.~\ref{fig7} for the nn QFS and in Fig.~\ref{fig8} 
 for the SST 
configurations. Again there is  an angle around which the contribution 
of that component is minimized. For the nn QFS it occurs around  
$\theta^{lab}_1 \approx 28^o$ and for the SST around 
 $\theta^{c.m.}_{plane} \approx 90^o$.

\section{Summary}
\label{summary}

We investigated the importance of the scattering amplitude components 
with the total 3N isospin $T=3/2$ 
in two 3N reactions. The inclusion of these components is required to 
 account for 
CIB effects of the NN interaction. The difference between np and 
 nn (pp) forces leads to a situation in which  also the matrix elements of the 
3NF between $T=3/2$ 
states contribute to the considered 3N reactions. The modern NN 
interactions, which describe  existing  pp and np data 
with high precision, provide pp and np t-matrices 
which differ up to $\approx 10 \%$. Such a magnitude of CIB requires 
that the isospin $T=3/2$ components are included 
in the calculation of the breakup reaction, especially for the
regions of the breakup phase-space close to the FSI condition. 
However, in order to account for 
all CIB effects it is sufficient to restrict the inclusion 
of $T=3/2$ to the $^1S_0$ 
 partial wave state 
only instead of doing it in  all $t=1$ states. For elastic scattering we found 
that the $T=3/2$ components can be neglected completely and all CIB effects are
accounted for by restricting oneself to total 3N isospin $T=1/2$ partial 
waves only and 
using the effective t-matrix generated with the $``2/3-1/3''$ rule 
$t_{\rm eff}=(2/3)t_{nn}+(1/3)t_{np}$. 
 These results allow one to reduce significantly the number of partial waves 
in time-consuming 3N 
calculations. This  is of particular  importance in view of the necessity 
to fix the parameters  of the  higher-order chiral 3NF components  
by fitting them to 3N scattering  observables. 

The presented results show that in 3N reactions the $T=3/2$ components 
 are overshadowed by the dominant 
$T=1/2$ contributions.  
It will be interesting to investigate reactions with three nucleons 
 in which only $T=3/2$ components contribute in the final state such as e.g. 
$^3H + \pi^- \to n + n + n$. That will allow one to study the properties and 
the importance of 3NFs in the $T=3/2$ states.

\begin{acknowledgements}
This work was performed by the LENPIC collaboration with support from 
  the Polish National Science Center 
 under Grant No. DEC-2013/10/M/ST2/00420 and 
 PRELUDIUM DEC-2013/11/N/ST2/03733, BMBF (contract
 No. 05P2015 - NUSTAR R$\&$D), and ERC Grant No. 307986 STRONGINT. 
 The numerical calculations 
have been performed on the supercomputer cluster of the JSC, J\"ulich, 
Germany.
\end{acknowledgements}

%
%

%
%

%
%

%
%

%
%

%
%

%
%

%
%

\begin{table}
\begin{tabular}{|c|c|c|c|c|c|c|c|c|}
\hline
 & & & & & & & & \\
 1 & 2 & 3 & 4 & 5 & 6 & 7 & 8 & 9 \\
 & & & & & & & & \\
\hline
  &  &  & & & & &  & \\
$E_{lab}$ & no CIB     & no CIB     & no CIB  & no CIB  &  $^1S_0$ CIB & $^1S_0$ CIB &    CIB &  CIB \cr
[MeV]  & $V_{123}=0$    & $V_{123}=0$   & $V_{123}=0$ & $V_{123}$ & $V_{123}=0$  & $V_{123}$ & $V_{123}=0$ & $V_{123}$ \cr
  & $^1S_0$ np   & $^1S_0$ nn & $t_{\rm eff}$ & $t_{\rm eff}$ &  &  &  & \cr
 & & & & & & & & \\
\hline
  & & & & & & & & \\
13.0 & 867.0 & 858.8 & 861.5 & 874.6  & 861.6 & 874.8 & 861.6 & 874.8   \\
  & & & & & & & & \\
\hline
 & & & & & & & & \\
65.0 & 163.4 & 159.4  & 160.7 & 169.3 & 160.7 & 169.4 & 160.7 & 169.3 \\
  & & & & & & & & \\
\hline
 & & & & & & &  & \\
135.0 & 77.21  & 75.91 & 76.34 & 80.92 & 76.34  & 80.93 &  76.34 &   80.91 \\
  & & & & & & & & \\
\hline
 & & & & & & & & \\
250.0 & 53.35  & 53.41 & 53.39 & 55.48 & 53.39 & 55.49 & 53.39 & 55.48 \\
 & & & & & & & & \\
\hline
\end{tabular}
\caption{The nd total cross section (in [mb]) at energies given in the first 
 column. Dynamical models related to the 
particular columns are: in 2nd, 3rd, 4th, and 5th no CIB in 
any $t=1$ states was 
assumed. In the $^1S_0$ state the t-matrix has been taken
 as $t_{np}$ and $t_{nn}$ 
for  the 2nd and 3rd, and as $t_{\rm eff}=(2/3)t_{nn}+(1/3)t_{np}$ for the 
 4th and 5th. 
In all other 
$t=1$ states $t_{\rm eff}$  was used. 
 Descriptions $V_{123}=0$ and $V_{123}$ means that the NN interaction 
was taken alone and combined with the 3NF, respectively. In the 
6th and 7th columns CIB was 
exactly taken into account by using in $^1S_0$ state both $t_{np}$ and $t_{nn}$ 
t-matrices and state with isospin $T=3/2$ was taken into account. 
 In all other $t=1$ states effective  t-matrix $t_{\rm eff}$ 
was used. In the 8th and 9th columns for all $t=1$ states CIB was 
 treated exactly by taking 
in addition to $T=1/2$ also $T=3/2$ states and the corresponding 
$t_{np}$ and $t_{nn}$ t-matrices.}
\label{table1}
\end{table}

%
%

\begin{table}
\begin{tabular}{|c|c|c|c|c|c|c|c|c|}
\hline
 & & & & & & & & \\
 1 & 2 & 3 & 4 & 5 & 6 & 7 & 8 & 9 \\
 & & & & & & & & \\
\hline
  &  &  & & & & &  & \\
$E_{lab}$ & no CIB     & no CIB     & no CIB  & no CIB  &  $^1S_0$ CIB & $^1S_0$ CIB &    CIB &  CIB \cr
[MeV]  & $V_{123}=0$    & $V_{123}=0$   & $V_{123}=0$ & $V_{123}$ & $V_{123}=0$  & $V_{123}$ & $V_{123}=0$ & $V_{123}$ \cr
  & $^1S_0$ np   & $^1S_0$ nn & $t_{\rm eff}$ & $t_{\rm eff}$ &  &  &  & \cr
 & & & & & & & & \\
\hline
 & & & & & & & & \\
13.0 & 699.8 & 699.3  & 699.4 & 711.8  & 699.5 & 712.0 & 699.5 &   712.0 \\
 & & & & & & & & \\
\hline
 & & & & & & & & \\
65.0 & 71.43 & 69.53 & 70.15 & 75.68 & 70.15 & 75.70 & 70.15 & 75.67  \\
 & & & & & & & & \\
\hline
 & & & & & & & & \\
135.0 & 20.80  & 20.31 & 20.46 & 22.63 & 20.46 & 22.64 & 20.46 & 22.62  \\
 & & & & & & & & \\
\hline
 & & & & & & & & \\
250.0 &  8.769 & 8.774 & 8.769 & 9.472 & 8.769 & 9.475 & 8.769  & 9.471 \\
 & & & & & & & & \\
\hline
\end{tabular}
\caption{The  nd elastic scattering total cross section (in [mb]). 
For the description of underlying dynamics see Table~\ref{table1}.}
\label{table2}
\end{table}

%
%

\begin{table}
\begin{tabular}{|c|c|c|c|c|c|c|c|c|}
\hline
 & & & & & & & & \\
 1 & 2 & 3 & 4 & 5 & 6 & 7 & 8 & 9 \\
 & & & & & & & & \\
\hline
  &  &  & & & & &  & \\
$E_{lab}$ & no CIB     & no CIB     & no CIB  & no CIB  &  $^1S_0$ CIB & $^1S_0$ CIB &    CIB &  CIB \cr
[MeV]  & $V_{123}=0$    & $V_{123}=0$   & $V_{123}=0$ & $V_{123}$ & $V_{123}=0$  & $V_{123}$ & $V_{123}=0$ & $V_{123}$ \cr
  & $^1S_0$ np   & $^1S_0$ nn & $t_{\rm eff}$ & $t_{\rm eff}$ &  &  &  & \cr
 & & & & & & & & \\
\hline
 & & & & & & & & \\
13.0 & 167.2 & 159.8  & 162.1 & 162.9 & 162.1 & 162.8 & 162.1 &   162.9 \\
 & & & & & & & & \\
\hline
 & & & & & & & & \\
65.0 & 91.92  & 89.89 & 90.58 & 93.69 & 90.58 & 93.69 & 90.58  & 93.61\\
 & & & & & & & & \\
\hline
 & & & & & & & & \\
135.0 & 56.42  & 55.60 & 55.88 & 58.29 & 55.88 & 58.29 & 55.88 & 58.28 \\
 & & & & & & & & \\
\hline
 & & & & & & & & \\
250.0 & 44.58  & 44.64 & 44.62 & 46.01 & 44.62 & 46.01 & 44.62  & 46.01 \\
 & & & & & & & & \\
\hline
\end{tabular}
\caption{The nd breakup total cross section (in [mb]). 
For the description of underlying dynamics see Table~\ref{table1}.}
\label{table3}
\end{table}

%
%


%

\end{document}